\documentclass[twocolumn,amsmath,amssymb]{revtex4}
\usepackage{amssymb}
\usepackage[dvips]{graphicx}
\begin{document}

\title{Strong-coupling superconductivity beyond BCS and the key pairing interaction in cuprate superconductors}
\author{A. S. Alexandrov}

\affiliation{Department of Physics, Loughborough University,
Loughborough LE11 3TU, United Kingdom, e-mail:
a.s.alexandrov@lboro.ac.uk\\}

\begin{abstract}
It has been now over 20 years since the discovery of the first high
temperature superconductor by Georg Bednorz and Alex M\"uller in
1986 and yet, despite intensive effort, no universally accepted
theory exists about the origin of high-temperature
superconductivity. A controversial issue on whether the
electron-phonon interaction (EPI) is crucial for high-temperature
superconductivity or weak and inessential has been one of the most
challenging problems of contemporary condensed matter physics. I
briefly review our recent theoretical results, which in conjunction
with a great number of experimental observations including isotope
effects, angle-resolved photoemission (ARPES), pump-probe and
tunnelling spectroscopies, normal state diamagnetism and magnetic
quantum oscillations provide the definite answer to this fundamental
question. The true origin of high-temperature superconductivity is
found in  a significant finite-range Fr\"ohlich EPI of
\emph{nonadibatic} polaronic carriers which is beyond the
conventional BCS-Migdal-Eliashberg  approximation.

\end{abstract}

\maketitle

After we have shown  \cite{ale0}--- unexpectedly
 for many
 researchers--- that   the BCS-Migdal-Eliashberg (BCS-ME) theory  breaks down already at the EPI coupling $\lambda \gtrsim 0.5$ for any
 adiabatic ratio $\hbar \omega_0/E_F$, the multi-polaron physics has gained particular
attention \cite{aledev}.  The  parameter $\lambda
\hbar\omega_0/E_F$, which is supposed to be  small in the BCS-ME
theory   becomes in fact large at $\lambda \gtrsim 0.5$ since the
electron band  narrows and the Fermi energy, $E_F$ turns out  below
the characteristic phonon energy, $\hbar \omega_0$. Nevertheless, as
noted in the unbiased comment by Jorge Hirsch \cite{hirsh}, in order
to explain the increasingly higher T$_c$s found in supposedly
'conventional' materials,  values of the electron-phonon coupling
constant $\lambda$ larger than $1$ have been used in the
conventional BCS-ME formalism.  This formalism completely ignores
the polaronic collapse of the bandwidth, but regrettably continues
to be  used by some researchers (irrespective of whether $\lambda$
is small or large) failing to explain high-T$_c$. As a result, many
researchers maintain that a repulsive electron-electron interaction
is responsible for pairing providing high transition temperatures,
$T_c$, without phonons in high-temperature superconductors.
\begin{figure}
\begin{center}
\includegraphics[angle=-90,width=0.60\textwidth]{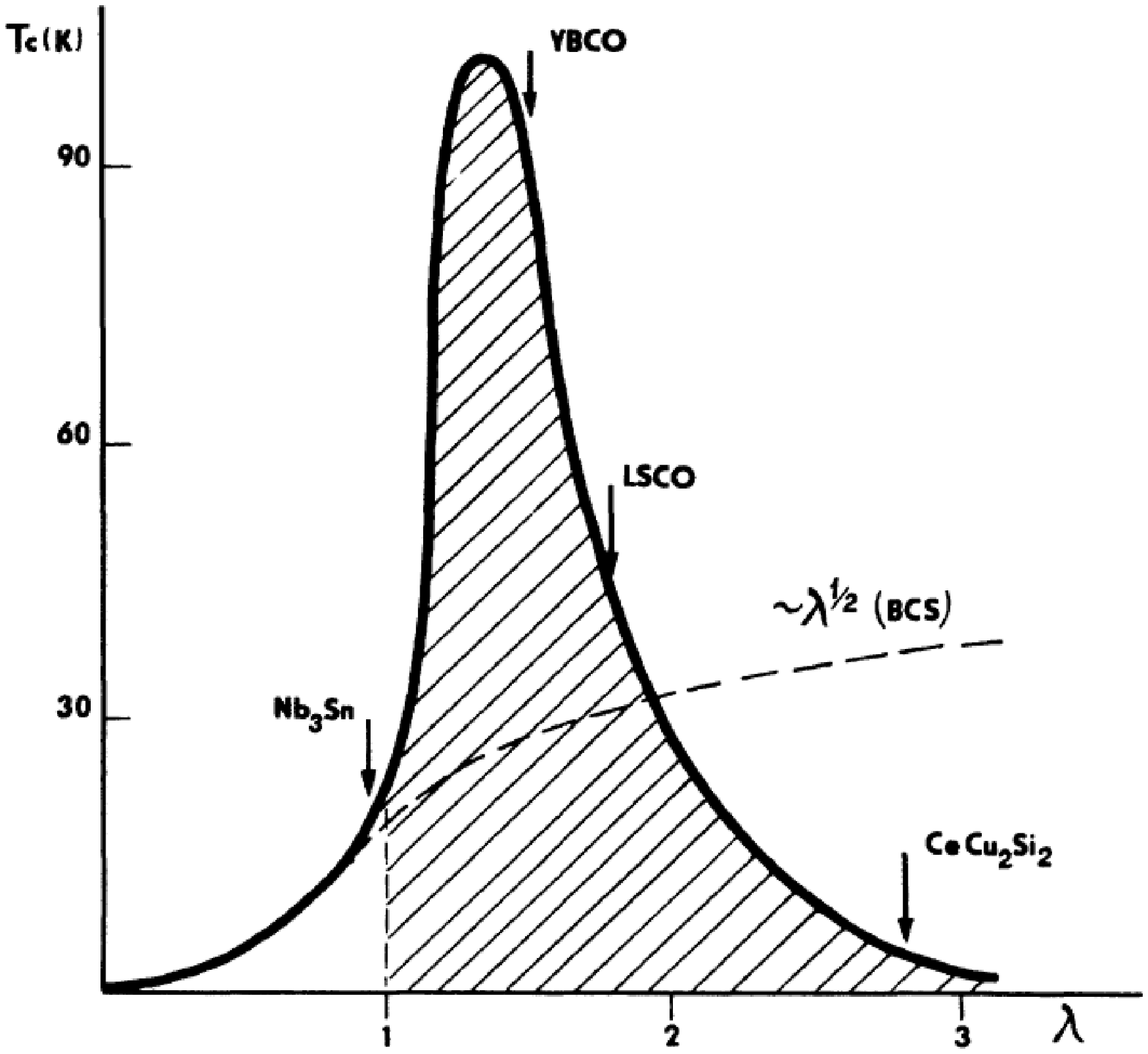}
\vskip -0.5mm \caption{ Breakdown of the BCS-ME approximation due to
the polaron bandwidth collapse at some critical
 coupling  $\lambda$ of the order of 1.  The dependence of T$_c$
versus $\lambda$ is shown in the polaron-bipolaron crossover region
compared with the BCS-ME dependence (dotted line). (Reproduced from
 Alexandrov A S 1988 \emph{ Phys. Rev. B} \textbf{38} 925,
(\copyright ) American Physical Society, 1988.)}
\end{center}
\end{figure}

Properly extending the BCS theory towards the strong interaction
between electrons and ion vibrations, a \emph{Bose liquid }of
tightly bound electron pairs surrounded by the lattice deformation
(i.e. of \emph{ small bipolarons}) was predicted \cite{aleran}.
Further prediction was that high temperature superconductivity
should exist in the crossover region of the electron-phonon
interaction strength from the BCS-like to bipolaronic
superconductivity \cite{ale0}, Fig.1. The strong enhancement of
T$_c$ in the crossover region from BCS-like polaronic to BEC-like
bipolaronic superconductivity is entirely due to a sharp increase of
the density of states in a narrow polaronic band \cite{ale0}, which
is missing in the so-called \emph{negative} Hubbard $U$ model.

We have recently quantified the EPI strength, the phonon-induced
electron-electron attraction, and the carrier mass renormalization
in layered superconductors at different doping using a continuum
approximation for the renormalized carrier energy spectrum and the
RPA dielectric response function \cite{alebra}. If, for instance we
start with a parent insulator as La$_{2}$CuO$_{4}$,  the magnitude
of the Fr\"{o}hlich EPI is unambiguously estimated using the static, $%
\epsilon _{s}$ and high-frequency, $\epsilon _{\infty }$ dielectric
constants. To assess its strength, one can apply an expression for
the polaron binding energy (polaronic level shift) $E_{p}$, which
depends only on the measured $\epsilon _{s}$ and $\epsilon _{\infty
}$,
 $E_{p}=(e^{2}/2\epsilon _{0}\kappa)\int_{BZ}d^{3}q/(2\pi
)^{3}q^{2}.$ Here, the integration goes over the Brillouin zone
(BZ), $\epsilon _{0}\approx 8.85\times 10^{-12}$ F/m is the vacuum
permittivity, and $\kappa =\epsilon _{s}\epsilon _{\infty
}/(\epsilon _{s}-\epsilon _{\infty })$. In the parent insulator, the
Fr\"{o}hlich interaction alone provides the binding energy of two
holes, $2E_{p}$, \emph{an order of magnitude} larger than any
magnetic interaction ($E_{p}=0.647$ eV in La$_{2}$CuO$_{4}$).

 Recent observations of the quantum magnetic
oscillations in some cuprate superconductors \cite{und}  are opening
up a possibility for a quantitative assessment of EPI in these and
related doped ionic lattices with the quasi two-dimensional (2D)
carrier energy spectrum. The oscillations revealed almost
cylindrical Fermi surfaces, enhanced effective masses of carriers
(ranging from $2m_{e}$ to $6m_{e}$) and the astonishingly low Fermi
energy, $E_F$, which is well below 40 meV in Y-Ba-Cu-O.
 Such low Fermi
energies  make the Migdal-Eliashberg (ME) adiabatic approach to EPI
 inapplicable in
these compounds since the characteristic oxygen vibration energy
(about $\hbar \omega _{0}=80$ meV) turns out larger than the carrier
kinetic energy. Since carriers in cuprates are in the non-adiabatic
(underdoped) or near-adiabatic (overdoped) regimes, $E_{F}\lesssim
\hbar \omega _{0}$, their energy spectrum renormalized by EPI and
the polaron-polaron interactions can be found with the familiar
small-polaron canonical transformation at \emph{any coupling
}$\lambda $ \cite{alekor}.

With doping the attraction and the polaron mass drop \cite{alebra}.
Nevertheless,  on-site  and  inter-site attractions induced by EPI
remain several times larger than the superexchange (magnetic)
interaction $J$ (about 100 meV) at any doping since the
non-adiabatic carriers cannot fully screen high-frequency electric
fields.  The polaron mass \cite{alebra} agrees quite well with the
experimental masses
measured in magnetic quantum oscillations experiments.  Hence the Fr\"{o}%
hlich EPI with high-frequency optical phonons turns out to be the
key pairing interaction in underdoped cuprates and remains the
essential player at overdoping. What is more surprising is that EPI
is clearly beyond the BCS-ME approximation since its magnitude is
larger than or comparable with the Fermi energy and the carriers are
in the non-adiabatic or near-adiabatic regimes. Since EPI is not
local in the nonadibatic electron system with poor screening it can
provide the d-wave symmetry of the pairing state \cite{aledwave}.

There are other independent pieces of evidence in favor of
(bi)polarons  and 3D BEC in cuprate superconductors. Most compelling
evidence for  (bi)polaronic carries in cuprate superconductors is
the  substantial isotope effect on the carrier mass \cite{alezhao}.
High resolution ARPES \cite{meevasana} provides another piece of
evidence for a strong electron-phonon interaction (EPI) in cuprates
and related oxides apparently with c-axis-polarised optical phonons.
These as well as recent pump-probe experiments \cite{dragan}
unambiguously show that the Fr\"{o}hlich EPI is important in those
highly polarizable ionic lattices.

Magnetotransport
 data strongly support preformed bosons in cuprates.
In particular,
 many high-magnetic-field studies revealed a non-BCS upward
curvature of the upper critical field $H_{c2}(T)$,  predicted for
the Bose-Einstein condensation of charged bosons in the magnetic
field \cite{aleH}. Nonlinear normal-state diamagnetism of quite a
few hole-doped cuprates has a profile characteristic of normal state
real-space composed bosons (i.e. bipolarons) \cite{aledia}, rather
than "preformed" Cooper pairs, vortex liquid and the
Kosterlitz-Thouless phase transition hypothesized by some authors.

Single polarons, \emph{localised }within an impurity band-tail,
coexist with bipolarons in  charge-transfer doped Mott-Hubbard
insulators. They account for  sharp \textquotedblleft
quasi-particle" peaks near $(\pi/2,\pi/2)$ of the Brillouin zone and
high-energy \textquotedblleft waterfall" effects observed with ARPES
in cuprate superconductors \cite{alekim}. This "band-tail" model
also accounts for two energy scales (superconducting and
pseudo-gaps) in ARPES \cite{alekim} and in the extrinsic and
intrinsic tunnelling \cite{alebin}.

All these and many other observations point to a crossover from the
bipolaronic to polaronic superconductivity \cite{ale0} in
high-temperature superconductors with doping.


\begin{thebibliography}{90}

\bibitem{ale0}   Alexandrov A S 1983   \emph{Zh. Fiz. Khim. } \textbf{57}
273  [1983 \emph{Russ. J. Phys. Chem.} \textbf{57} 167]
\bibitem{aledev} Alexandrov A S and Devreese J T 2009 Advances in
Polaron Physics (Heidelberg-Dordrecht-London-New York: Springer)
\bibitem{hirsh} Hirsh J E 2009  \emph{Phys. Scr.} \textbf{80}  035702
\bibitem{aleran}   Alexandrov A S and Ranninger J 1981  \emph{Phys.
Rev. B}  \textbf{23} 1796; 1981 \emph{Phys. Rev. B}\textbf{ 24} 1164
\bibitem{alebra} Alexandrov A S and Bratkovsky A M 2010 \emph{Phys. Rev.
Lett.} \textbf{105}  226408
\bibitem{und} Doiron-Leyraud N \emph{et al} 2007  \emph{Nature (London)} \textbf{447}%
 565
\bibitem{alekor} Alexandrov A S and
Kornilovitch P E 1999  \emph{Phys. Rev. Lett.} \textbf{82} 807
\bibitem{aledwave}  Alexandrov A S 2008
\emph{ Phys. Rev. B} \textbf{77} 094502
\bibitem{alezhao}  Alexandrov A S and Zhao G M 2009
\emph{Phys. Rev. B }\textbf{80} 136501
\bibitem{meevasana}  Meevasana W \emph{et al} 2010  \emph{New J. Phys.} \textbf{12} 023004
\bibitem{dragan}  Gadermaier C \emph{et al} 2010 \emph{Phys. Rev.
Lett.} \textbf{105}  257001
\bibitem{aleH}  Alexandrov A S 1993
\emph{Phys. Rev. B} \textbf{48} 10571
\bibitem{aledia}  Alexandrov A S 2006   \emph{Phys. Rev. Lett.}
\textbf{96} 147003; 2010  \emph{J. Phys.: Condens. Matter}
\textbf{22} 426004
\bibitem{alekim}
 Alexandrov A S and  Reynolds K  \emph{Phys. Rev. B}  \textbf{76} 132506
\bibitem{alebin}  Alexandrov A S and  Beanland J 2010
\emph{Phys. Rev. Lett.} \textbf{104} 026401







\end{thebibliography}
\end{document}